# Extraordinary water adsorption characteristics of graphene oxide

B. Lian[1], S. De Luca[2], Y. You[3], S. Alwarappan[4], M. Yoshimura[5], V. Sahajwalla[3], S.C. Smith[2], G. Leslie[1], R.K. Joshi[3]

[1]UNESCO Centre for Membrane Science and Technology
School of Chemical Engineering, University of New South Wales Sydney, Australia.
[2]Integrated Materials Design Centre,
School of Chemical Engineering, University of New South Wales Sydney, Australia.
[3]Centre for Sustainable Materials Research and Technology (SMaRT),
School of Materials Science and Engineering, University of New South Wales, Sydney Australia
[4]CSIR- Central Electrochemical Research Institute, Karaikudi 630003, Tamilnadu, India.
[5]Surface Science Laboratory, Toyota Technological Institute, Nagoya, Japan.

*Corresponding author's email: r.joshi@unsw.edu.au*

**The laminated structure of graphene oxide (GO) confers unique interactions with water molecules which may be utilised in a range of applications that require materials with tuneable hygroscopic properties. Precise roles of the expandable interlayer spacing and functional groups in GO laminates are not fully understood till date. Herein, we report experimental and theoretical study on the adsorption and desorption behaviour of water in GO laminates as a function of relative pressure. We have observed that GO imparts excellent water uptake capacity of up to 0.58 gram of water per gram of GO (g g$^{-1}$), which is much higher than silica gel a conventional desiccant material. More interestingly, the adsorption and desorption kinetics of GO is one order of magnitude higher than silica gel. The observed extraordinary adsorption/desorption rate can be attributed to the high capillary pressure in GO laminates as well as micro meter sized tunnel like wrinkles located at the surface.**

Nanoporous materials with high surface area and large pore volume are often employed as desiccant materials [1-3]. The inhomogeneous 3-D porous materials such as silica gel, zeolite and metal organic frameworks are the widely used desiccant materials [4-7]. However, issues such as large pore size distribution, low surface area to pore volume ratio, low hydrophilicity and poor hydrothermal stability associated with the aforementioned materials offer limitations for wide applicability [1, 8-11]. Recent studies on the interaction of water with graphene oxide laminates have demonstrated the possibility of utilizing its excellent hydrophilicity for numerous applications[12-14]. Being a 2-D porous material, graphene oxide not only possesses uniform pore size but also finds diverse functionalization potential, ultra-fast water transport mechanism and expandable interlayer spacing[15-19]. All these features and the formation of hydrogen bond network of water micro-cluster in the confined GO laminates significantly affects water molecule's diffusion rate, and potential energy at absorbed state [20-22]. The strong interaction between GO and water makes it a potential candidate for desiccation application. Herein, we have studied the water adsorption capacity and kinetics of GO extensively and reported in this manuscript.

Initially, we investigated the water vapour uptake of GO, silica gel, graphite and reduced graphene oxide (rGO) at different relative pressure $P\ P_0^{-1}$ (where $P_0$ represents saturation pressure) from 0.1 to 0.9. Results are summarised in Figure 1. Both graphite and rGO show insignificant adsorption ability with values less than 0.05 g g$^{-1}$ water uptake (1a). Although, the pore size of graphite (3.4 Å) and rGO (3.7 Å)[23] should be enough to accommodate water molecule with size of 2.4 Å, their hydrophobic characteristics restricts the entry of water molecules into the pores of these materials [24]. On the other hand, due to the larger proportion of hydrophilic functional groups, membrane like GO prepared by vacuum filtration, exhibits excellent adsorption capacity, which is at least two times higher than the conventional desiccant material such as silica gel (pore size 2-6 nm) across the tested range of relative pressure. The water uptake of GO is 0.13 g g$^{-1}$ at a low relative pressure ($P\ P_0^{-1}$) of 0.1, which reaches to 0.58 g g$^{-1}$ at relative pressure of = 0.9.

In our study the trend of the adsorption isotherm is in excellent agreement with type- II isotherm classified by International Union of Pure and Applied Chemistry (IUPAC) [25]. However, type II is often used to describe the adsorption behaviour of non-porous or macro-porous hydrophobic materials, while type I and IV are for microporous and mesoporous materials respectively. This is contradictory to the existence of GO microporous structure and its hydrophilic nature. On the other hand, the traditional porous materials characterized by IUPAC type have rigid porous structure, while the interlayer spacing of GO can be varied under wet conditions. Figure 1 b shows the optical image of GO laminates in the membrane form at different hydration states. GO membranes dried at 80°C for 10 mins are folded whereas the wet membranes at a relative pressure of $P\ P_0^{-1}$ = 0.6 remain flat. Here the folding of GO membrane can be attributed to the reduction of pore size which is interlayer d-spacing for GO laminates. This can be further confirmed by the X-ray diffraction (XRD) measurements as the d-spacing of GO sample changed from 6.5 Å at dry condition to 11.3 Å at 0.9 $P P_0^{-1}$ (Figure 1c). At relative pressures below 0.3, the d-spacing of GO was only enhanced by 1.8 Å as shown in XRD plots, which is less than the size of a water molecule (~2.4 Å). This suggests that the adsorption is mainly due to the strong interaction between water-GO surfaces that corresponds to type I IUPAC adsorption isotherm reported for microporous materials. However, when the relative pressure increases above 0.6, d-spacing was enhanced by 2.2 Å at $P P_0^{-1}$ = 0.6 to 4.8 Å at $P P_0^{-1}$ = 0.9 respectively resulting in a proportional increase in the water uptake from 0.28 g g$^{-1}$ to 0.58 g g$^{-1}$. This evidenced the multilayer water formation in GO laminates by capillary condensation shown typically by the mesoporous material in accordance with IUPAC type IV adsorption isotherm (Figure 1a). Thus, we believe that the adsorption isotherm of GO observed in the present work is a combination of the IUPAC type I and type IV.

We performed molecular dynamics (MD) simulations to probe more insights onto water behaviour within GO laminates at different relative pressure conditions. MD simulation accurately predicted the water adsorption capacity of GO with less than 7% deviation from the experiment (Figure 1a). The water density profile in GO laminates obtained from the MD simulation were then plotted both along the thickness direction z (Figure 1d and 1d') and parallel to GO basal plain (Figure 1e). As depicted in Figure 1d, over 90% of water molecule positioned at approximately 0.34nm apart from the carbon pristine plain at $P P_0^{-1}$ = 0.3. This is due to the Van der Waals force between water and pristine carbon to maintain low potential energy, and it shows the significance of surface water interaction at low relative pressure demonstrated by the adsorption isotherm curve. Water density distribution profile along the GO plain suggests that the water molecules were closely packed around GO functional group at $P P_0^{-1}$=0.3 (Figure 1e, top right) whereas a spread water molecule distribution was found for higher relative pressure (Figure 1e, bottom). The average distance between water molecule and GO functional group was calculated as 0.32 nm at 0.3 $P P_0^{-1}$ and 0.47 nm at 0.9 $P P_0^{-1}$. At $P P_0^{-1}$ = 0.9, a distinct bimodal distribution was noticed for water with two peaks positioned 0.34 nm apart from the GO pristine plain, indicating the strong interaction between GO and water at high relative pressure. However, about 43% of water molecules positioned away from the low potential energy free spots where water-water interaction dominates. This proves the capillary condensation effect inside GO laminates at high relative pressure.

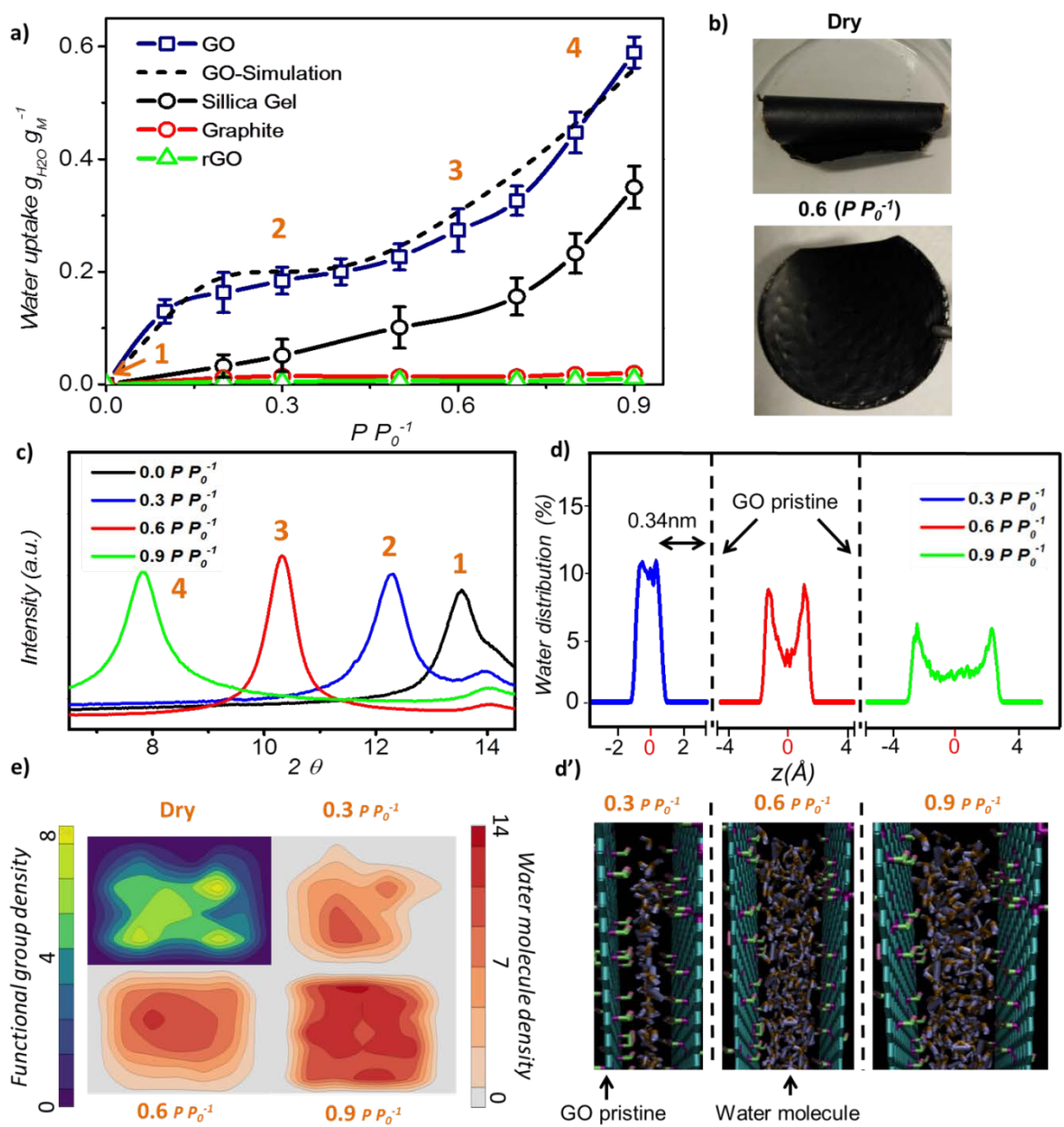

**Figure 1** *Water adsorption of graphene oxide.* a) Adsorption isotherm of GO, silica gel, graphite and rGO at 25 °C. b) photograph of GO laminates (top-dried at 80°C and bottom-saturated at $P P_0^{-1}$=0.6. c) XRD patterns of GO laminates under different conditions. d) MD simulated water density profile across the GO laminate at different relative pressure where z=0 represents centre of two GO plains and d' shows the water molecule position across two GO plains. e) Density profile of water molecules and GO functional groups parallel to the GO plain.

At this stage, free space between GO sheets determines the adsorption capacity. This is also proved by the additional MD simulations; where more defects or less functional group will increase the adsorption capacity of GO (Figure s2). In between these two stages, as the relative pressure increases from 0.3 to 0.6, water molecules start to occupy and as a result GO capillary expands. The slower increase rate in the water uptake at this stage is due to the high potential mean force between GO laminates that quickly attains the equilibrium with the environmental water vapour pressure[26, 27].

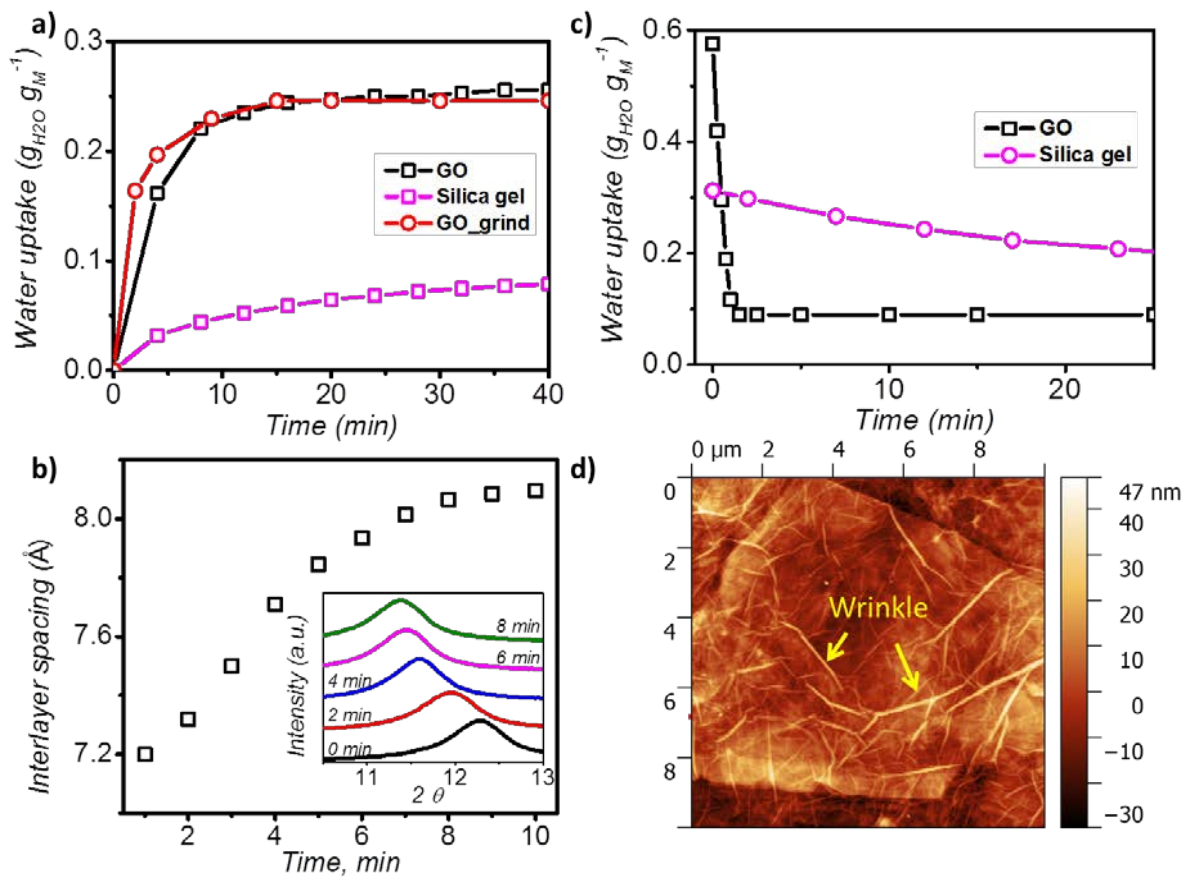

**Figure 2:** *Adsorption/desorption kinetic of GO. a) Water adsorption rate measured at 0.6 $P P_0^{-1}$ for 50 minutes. b) Variation of interlayer spacing of GO laminates with time; inset shows the XRD plots during the adsorption process of GO (in the range $2\theta$ = 6° to 16°). c) Water desorption rate at 40°C and 0.2 $P P_0^{-1}$ (the GO sample was saturated at $P P_0^{-1}$ = 0.9 for overnight prior to the water desorption). d) AFM images of GO showing 20 nm tunnel like wrinkles.*

In order to understand the kinetics of GO water adsorption and desorption, we measured the weight change of GO at 25°C (ambient condition) for adsorption rate, and at 40°C for desorption rate. The adsorption rate of GO was found to be 0.04 g $g^{-1}$ per minute, which is about five times faster than that of silica gel (Figure 2a). It was interesting to note that GO reached its maximum water adsorption capacity in 40 minutes from which the initial ~88% was achieved in just less than 10 minutes. We also studied the water adsorption capacity of grinded GO (average particle size ~0.5mm) and observed a higher adsorption rate which can be attributed to the higher exposed GO surface area. However, the grinding process affects the integrity of the GO capillaries which results into slight decrease in the adsorption capacity by 0.01 g g$^{-1}$.

XRD was further utilised to analyse the he expansion of d-spacing during GO water adsorption. In our specific XRD measurements, 10 individual scans for $2\theta = 6^{o}$ to $16^{o}$ were carried out continuously for 10 minutes during the water adsorption of GO with changing the relative pressure from 0.3 to 0.53 (Figure 2b). We observed expansion of d-spacing from 7.2 Å to 7.9 Å with a constant increase rate of 0.1Å min$^{-1}$ in the first 7 minutes of water adsorption, before reaching the equilibrium d value of 8.1Å for GO. It is also interesting to note that during the water adsorption process, XRD always shows one distinct peak shift rather than the exchange of two peaks. This confirms that the expansion of GO laminate is a collective movement of all laminates across the complete GO membrane.

During desorption experiments, we noticed that about 80% of water (by weight) desorbed at low regeneration temperature of 40°C with a very fast desorption rate of 0.46 g g$^{-1}$ per minute (Figure 2c). The fast desorption rate of GO can be due to combined effect of relatively higher temperature and low humidity. The moderate temperature of 40°C can accelerate the water molecule's diffusion rate and the low relative pressure creates larger pressure difference to further push water out of the GO capillaries to the environment. GO also showed a consistent lower water uptake than silica gel at tested regeneration temperature (40 to 100°C) (Figure s3), and exhibited excellent stability (Figure s4), proving it an ideal material for desiccant process.

In order to understand the rapid water adsorption and desorption rate of GO, MD simulation for GO water adsorption was performed. The simulation results for water adsorption kinetics confirmed that all water molecules are absorbed into by the capillaries inside the GO laminates within the first 1.5 ns and no water molecule coming out for 20 ns (Figure s5). This indicates the existence of a very strong capillary pressure in the GO even at low water uptake. It also shows that the rate of water adsorption is not limited by the speed of water entering into GO open pore, but controlled by the rate of water transport inside GO laminates. A common term that represents the molecular mobility in nanoporous structure in MD simulations is the self-diffusivity D. The simulated value of D for water molecule in GO at $P\ P_0^{-1}$ = 0.6 was equal to 0.131 × 10$^{-9}$ m$^2$ s$^{-1}$ (Figure s6), which is in agreement with value reported by Jiao et al., (~0.147× 10$^{-9}$ m$^2$ s$^{-1}$) and Devanathan et al., (~0.15 × 10$^{-9}$ m$^2$ s$^{-1}$)[28,29]. In comparison with the reported diffusivity of water in silica gel (in the range 0.28 × 10$^{-9}$ m$^2$ s$^{-1}$ to 1.5 × 10$^{-9}$ m$^2$ s$^{-1}$)[30] the simulated value D for water in GO laminates suggests low mobility based self-diffusion which is in contrary to our experimental results. Further experimental based analysis of the process is required to describe the fast water transport in GO.

Analysis of surface morphology using atomic force microscopy (AFM) further helps understand the rapid water adsorption and desorption of GO. We observed wrinkle (micrometre size) like structures with average height of 20 nm and up to 1.5 µm in size on GO surface (Figure 2d and Figure s7). Such winkles may act as tunnels for water transport which allow rapid water diffusion. Water transport can be facilitated by these wrinkles to distribute through GO before get adsorbed into GO laminates.

Furthermore, the adsorption isotherm (Figure 3) at 25°C and 40°C for GO membrane was calculated using the Clausius-Clapeyron relation.

$$\Delta h_{isos} = R \left( \frac{d(\ln P)}{d\left(-\frac{1}{T}\right)} \right)_\omega \tag{1}$$

Where $\Delta h_{isos}$, R, P, T, and ω represent the isosteric enthalpy of adsorption, universal gas constant, pressure, temperature, and water uptake, respectively. The isosteric enthalpy of water adsorption is 30% higher than that of other desiccant materials[32]. The simulated enthalpy change during adsorption was equal to -4062.3kJ.kg$^{-1}$, and it is in excellent agreement with the experimental data which further validates our MD module. Such a property is good for the applications such as heat pump.

Hydrogen bond analysis based on MD simulation confirmed that the interaction amongst water molecules is stronger than the interaction between water molecules and GO (Figure 3b). The ratio of the functional groups that forms hydrogen bond with water was found to increase from 0.27 to 0.37 with increasing humidity (from 0.15 to 0.75 $P\,P_0^{-1}$). However, saturation occurs at conditions above $P\,P_0^{-1}$ = 0.75. Under such conditions, the amount of functional groups that is expected to form hydrogen bond with water remains fixed. In contrast, the hydrogen bond between water molecules continuously increases from 1.9 to 2.4 hydrogen bonds per water molecule. This is expected, as the increase in the humidity expands the d spacing of GO laminates and this enhances the number of water molecules exposed to each other leads to higher number of hydrogen bonds per water molecule. However, after reaching a relative pressure ($P\,P_0^{-1}$) of 0.6, the hydrogen bonds saturates at a value of 2 to 3 hydrogen bonds per water. Below a relative pressure ($P\,P_0^{-1}$) of 0.3, the simulated average number (~2) of hydrogen bonds is lower than reported 2.3 hydrogen bonds per water molecule for liquid water[33]. This indicates that water exists in an intermediate sate between liquid and vapour.

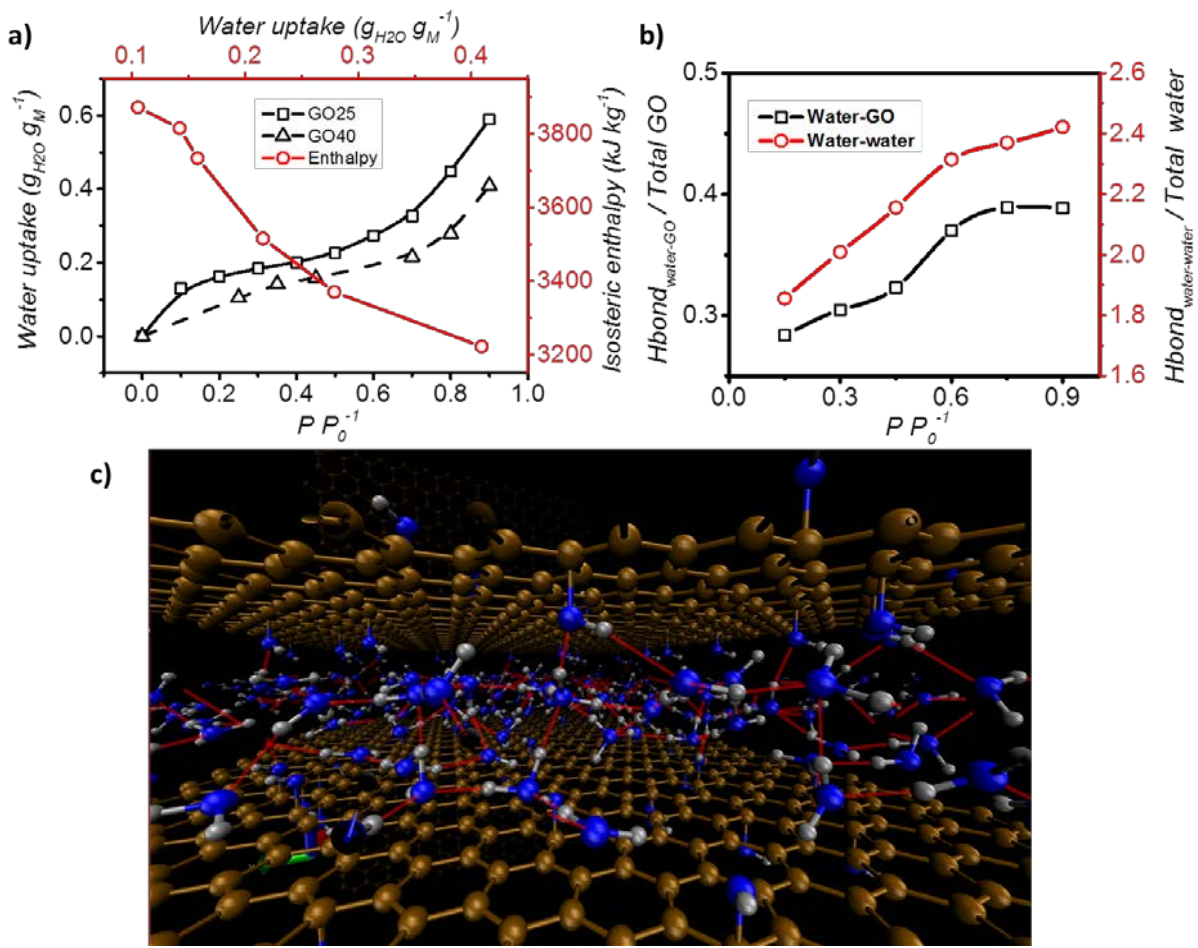

**Figure 3:** *Adsorption enthalpy and hydrogen bonding network.* a) *Adsorption isotherm of GO measured at both 25°C and 40°C, and the isosteric enthalpy of adsorption calculated based*

on the Clausius-Clapeyron relation. b) Representation of the interaction between water-GO functional group (Y axis-left) and water-water (Y axis -right) at different relative pressure. c) Schematic of the simulated GO-water configuration and hydrogen bond network at 0.6 $P P_0^{-1}$. C, O and H are shown in brown, blue and grey and hydrogen bonds are represented by red line (picture obtained with Visual Molecular Dynamics, VMD [31]).

In conclusion, GO laminates demonstrate remarkable water adsorption characteristics. The high water uptake capacity of GO is due to its expandable 2D porous laminated structure, and the fast water adsorption/desorption ability of GO can be attributed to the existence of wrinkle like water tunnel. Comparing with the commonly used desiccant material (silica gel) GO has higher water uptake capacity, significantly faster adsorption and desorption rate and higher adsorption enthalpy. These characteristics make it an advanced material for desiccant application as well as for heat pump processes.

# Extraordinary water adsorption characteristics of graphene oxide

B. Lian[1], S. De Luca[2], Y. You[3], S. Alwarappan[4], M. Yoshimura[5], V. Sahajwalla[3], S.C. Smith[2], G. Leslie[1], R.K. Joshi[3]

[1]UNESCO Centre for Membrane Science and Technology
School of Chemical Engineering, University of New South Wales Sydney, Australia.
[2]Integrated Materials Design Centre,
School of Chemical Engineering, University of New South Wales Sydney, Australia.
[3]Centre for Sustainable Materials Research and Technology (SMaRT),
School of Materials Science and Engineering, University of New South Wales, Sydney Australia
[4]CSIR- Central Electrochemical Research Institute, Karaikudi 630003, Tamilnadu, India.
[5]Surface Science Laboratory, Toyota Technological Institute, Nagoya, Japan.

*Corresponding author's email: r.joshi@unsw.edu.au*

**Methods:**

**GO membrane preparation:**

Graphene oxide was prepared using Hummer's method[1]. Briefly, 1.0 g of sodium nitrite and 2.0 g of graphite flakes with the size of 0.5 μm (from Sigma-Aldrich) is stirred with 48 mL of Con. $H_2SO_4$ in an ice bath. 4.6 g of $KMnO_4$ was then slowly added to the suspension to maintain temperature of the mixture below 15°C. The mixture was then stirred at room temperature for 30 mins, before being diluted using 100 mL Milli-Q water. The reaction vessel was kept at 98°C for half an hour and then 100 mL of 2% hydrogen peroxide solution was added. The solid in the suspension then gets separated and washed using 1.5 L of Milli-Q water  GO membrane were prepared by vacuum filtration of the resulting GO suspension through a 0.2 μm Polyvinylidene fluoride membrane . rGO was prepared by soaking GO membrane in 10% hydroiodic acid  for 4 hours. The resulted sample showed bright reflection of light and a hydrophobic feature.

**Adsorption of water vapour in GO:**

Adsorption equilibrium experiments were performed at atmospheric pressure in an Environmental chamber. The relative pressure was controlled to ±3% and the temperature could be controlled to an accuracy of ±1 K. A microelectronic balance with an accuracy of 0.0001 g was used to measure the sample weight. The dry mass of every prepared adsorbent placed on a Petri-dish was 0.05 g. The samples were pre-heated for 10 minutes and 30 minutes for GO and silica gel respectively at 80°C to discharge excess water prior testing at different temperatures and humidity. In order to measure the adsorption isotherm, all samples were kept in the environment chamber for at least 40 minutes. The sample weight was constantly inspected until no more weight change was noticed to ensure that the sample reach the adsorption equilibrium. Water adsorption rate was measured by recording the weight change of samples at 25°C and $P P_0^{-1}$ = 0.53 after being dried at 80°C. The samples were then placed in a desiccator for cooling prior to the weight measurement to illuminate heat effect on mass balance.

**Simulation methods:**

Classical molecular dynamics (MD) simulations were performed using the large-scale atomic/molecular massively parallel simulator (LAMMPS)[2]. The all-atom optimized potentials for liquid simulations (OPLS-AA)[3] were used for GO, which can capture essential many-body terms in interatomic interactions, including bond stretching, bond angle bending, nonbonding van der Waals and electrostatic interactions[4]. Three planar GO sheet (5*5 nm) with functional groups randomly seeded on the plane was created using MOLTEMPLATE [5]. Four graphene walls were set at the edge of upper and lower GO sheet to ensure water absorbed only in the two GO laminates. SPC/E model[6] was selected for water molecule with SHAKE algorithm[7] for constraining bond and angle. The interaction between water molecules and graphene or GO sheets includes both van der Waals and electrostatic terms. The former choice is described by the 12−6 Lennard−Jones potential $V_{LJ} = 4\epsilon[(\sigma/r)^{12} − (\sigma/r)^6]$ as a function of the interatomic distance r, with interaction parameters between water molecules and carbon atoms and functional group of GO are $\epsilon$ = 0.09369 kcal mol$^{-1}$, $\sigma$ = 3.19 Å and $\epsilon$ = 0.1553 kcal mol$^{-1}$, $\sigma$ = 3.17 Å respectively[8]. The partial charge of hydrogen, carbon and oxygen from GO was set to be 0.3294e, 0.1966e and -0.526e respectively[9]. The van der Waals forces are truncated at a distance of 1nm with a constant shift in the energy over the whole range to remove the discontinuity. The long-range Coulombic interactions were computed using the particle–particle particle-mesh algorithm (PPPM)[10].

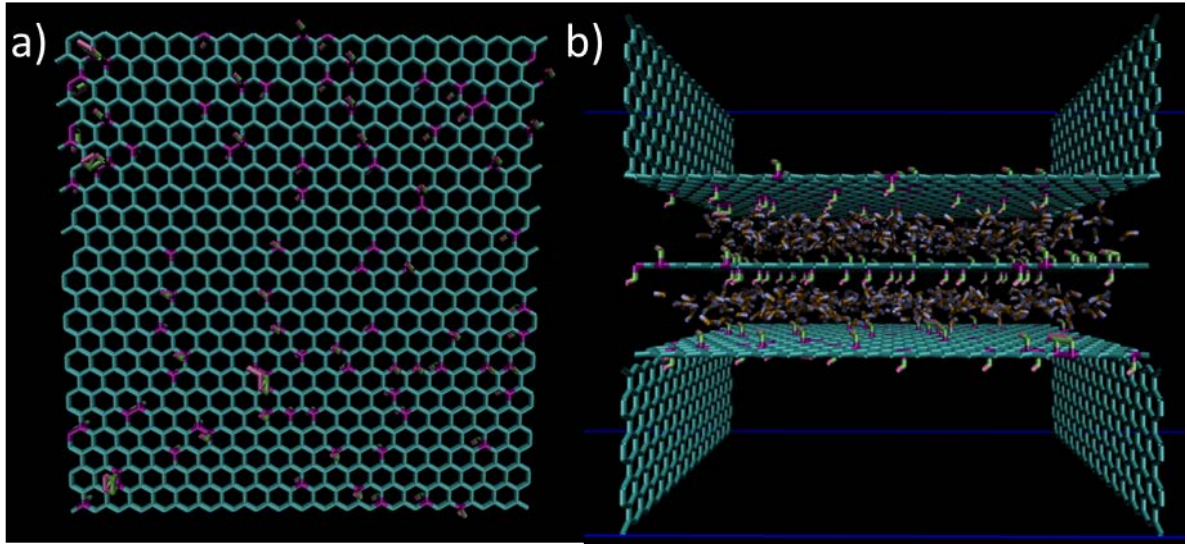

**Figure s1:** *MD simulation geometry with a) single GO flake at 10% oxidation level, functional group facing both side of the flake, b) 3 GO flake forming and water molecule in the two GO laminates. Four graphene walls were the constructed at the edge of the GO flake to prevent water molecule trapped at below and above the GO laminates in the simulation box.*

A simulation box of 5×5×4.5 nm was employed with periodic boundary condition applied at all directions (Figure s1). Detailed simulation geometry is shown in the supplementary information (Table s1). The simulation time step of 1 fs was selected for the integration of equation of motion. The overall system was minimized for 1000 time steps prior simulation and was electro-neutral to guarantee the convergence of the Ewald sum.

In order to predict the water uptake at different humid conditions, the simulations are carried out in the NVT ensemble with the Nosé–Hoover thermostat at 300K. Preselected interlayer spacing (based on XRD results) was used at different RH. The amount of water kept inside the GO laminate at the beginning of the simulation were continuously varied until no water molecule come out of the GO sheet until 10ns. 10% functional group as well as no defects was selected as default, while the variation of these two values was simulated under $P P_0^{-1}$ = 0.6.

The self-diffusion coefficient of water was calculated from the trajectory of water atom inside GO laminates at equilibrium state by using Einstein's correlation function between atomic position and diffusivity $D = \lim_{t \to \infty} (|r(t) - r(0)|)/2d_i t$, where r is the position of the atom, t is simulation time step and di is the dimensions of space for water movement. Due

to the 2D porous natural of GO, water movement in the normal direction is limited compare to the in plane motion. So the di = 2 was considered in this simulation[11]. The diffusivity was calculated using results from 0.5 ns after the simulation system reach equilibrium to ensure an accurate estimation of the diffusion constant. The adsorption enthalpy was calculated by comparing the total energy of water at a constant relative pressure and at the equilibrium state (when absorbed in GO laminates).

**Additional results**

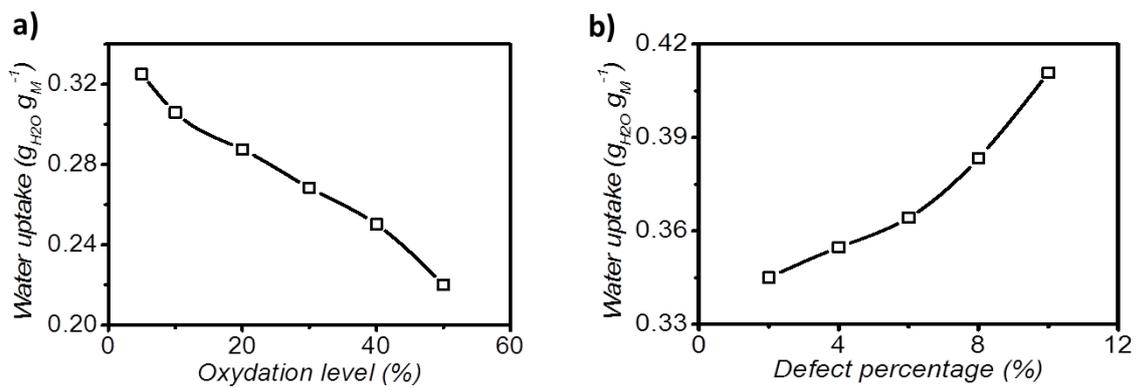

Figure s2: MD simulation predicted water uptake for modified GO membrane with a) different oxidation level, b) defect percentage

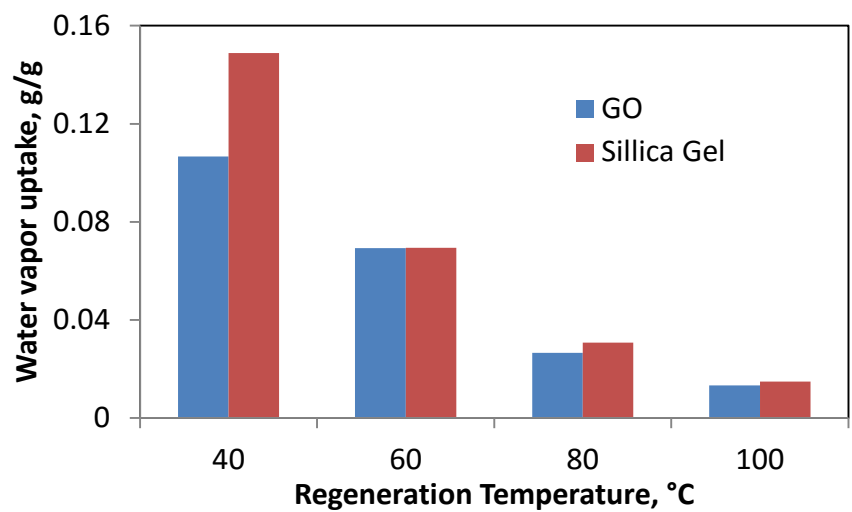

Figure s3: GO and silica gel's water uptake at different desorption temperatures

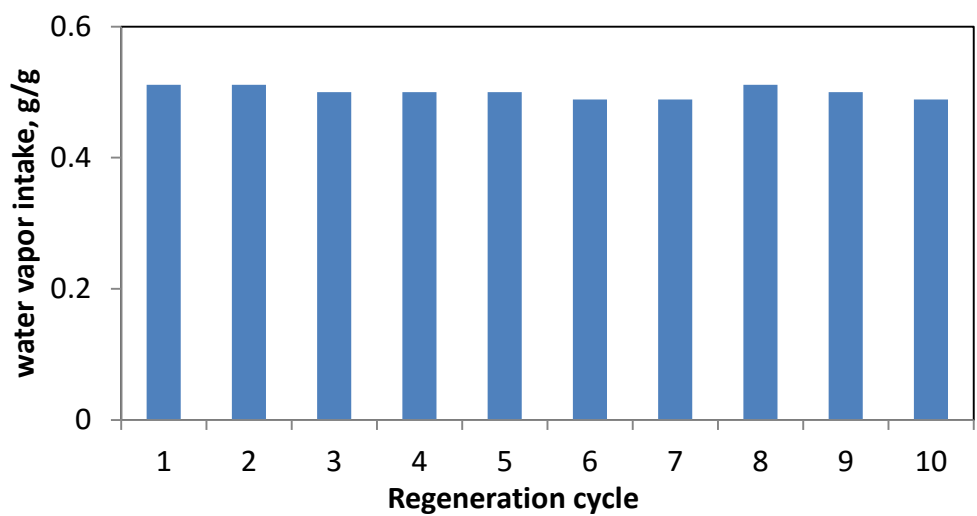

**Figure s4:** GO water adsorption capacity over 10 cycles of regeneration.

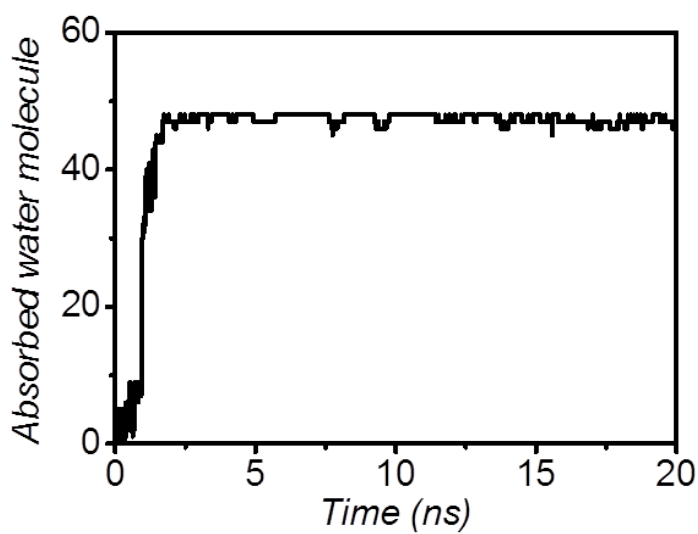

**Figure s5:** The number of water molecules (over 48 total water molecules) absorbed by GO as a function of time. The simulation was set at 60% relative humidity with a d-spacing of 8.6 Å for GO membrane.

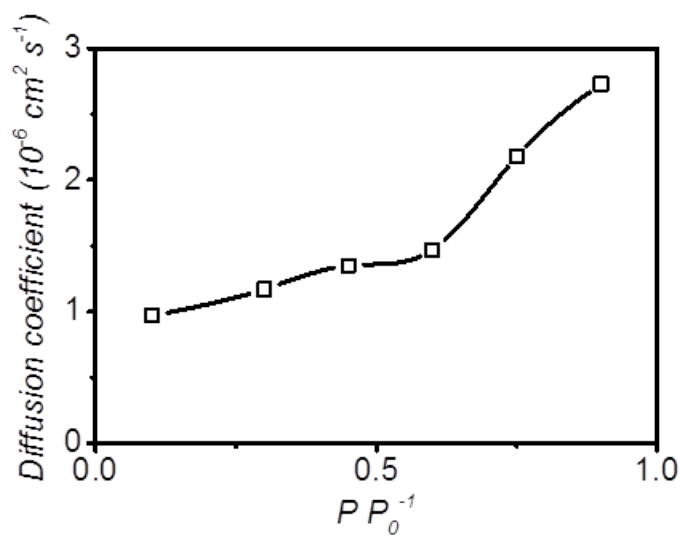

Figure s6: Simulated water molecule diffusion coefficient in GO at different humidity.

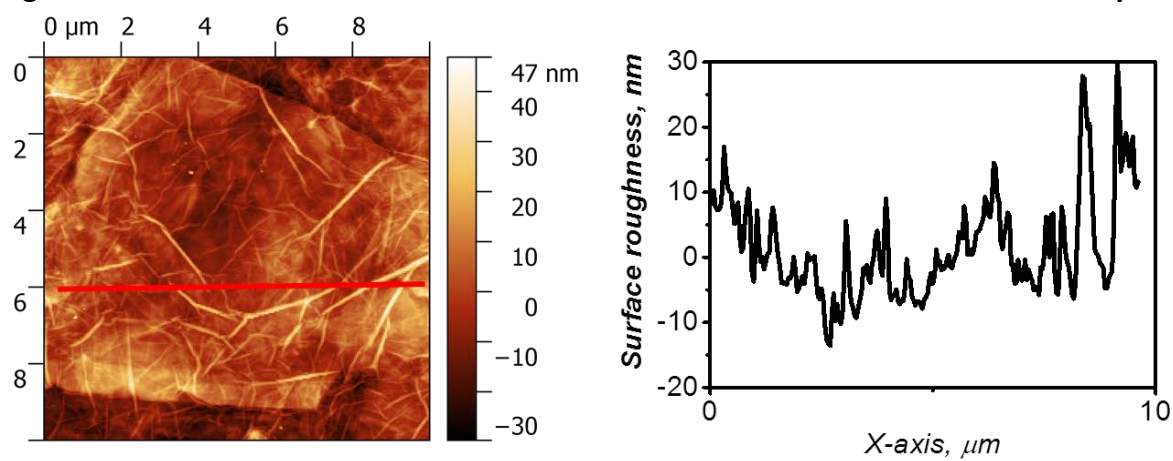

Figure s7: Surface roughness of GO at red line shown in AFM image (figure 2d main text)

Table s1: Summary of MD simulation for GO water adsorption at different relative pressure

| Description | Value | | |
|---|---|---|---|
| No. Flake | 3 | | |
| GO flack dimension | 5nm*5nm | | |
| Total carbon in GO | 2880 | | |
| Total functional group in GO | 288 | | |
| Relative pressure, $P P_0^{-1}$ | 0.15 | 0.3 | 0.45 |
| D spacing of GO, nm | 0.69 | 0.76 | 0.8 |
| No. Water molecule absorbed | 390 | 507 | 567 |
| $P/P_0$ | 0.6 | 0.75 | 0.9 |
| D spacing of GO, nm | 0.86 | 0.96 | 11.3 |
| No. Water molecule absorbed | 672 | 860 | 1268 |